\documentclass[12pt]{article}
\begin{document}

\author{C. Bizdadea\thanks{%
e-mail address: bizdadea@central.ucv.ro}, M. T. Miaut\u {a}, S. O. Saliu%
\thanks{%
e-mail address: osaliu@central.ucv.ro} \\
Faculty of Physics, University of Craiova\\
13 A. I. Cuza Str., Craiova RO-1100, Romania}
\title{Nonabelian interactions from Hamiltonian BRST cohomology }
\maketitle

\begin{abstract}
Consistent Hamiltonian couplings between a set of vector fields and a system
of matter fields are derived by means of BRST cohomological techniques.

\end{abstract}

\section{Introduction}

The cohomological approach to the Lagrangian BRST symmetry \cite{1}--\cite{5}
stimulated the incorporation of new aspects within the cohomological BRST
setting, like, for instance, a treatment of consistent interactions among
fields with gauge freedom with the preservation of the number of gauge
symmetries \cite{6}--\cite{10} from the perspective of the deformation of
the solution to the master equation \cite{11} with the help of the local
BRST cohomology \cite{13}--\cite{16a}. This procedure was proved to be an
efficient deformation technique for many models of interest, like
Chern-Simons models, Yang-Mills theories, the Chapline-Manton model, $p$%
-forms and chiral $p$-forms, Einstein's gravity theory, four- and
eleven-dimensional supergravity, or BF models \cite{11}, \cite{17}--\cite
{30b}.

In the meantime, the Hamiltonian version of BRST formalism \cite{5}, \cite
{31}--\cite{35} presents many useful and attractive features, like the
implementation of the BRST symmetry in quantum mechanics \cite{5} (Chapter
14), examination of anomalies \cite{36}, computation of local BRST
cohomology \cite{37}, and also the explanation of the relationship with
canonical quantization methods \cite{38}. Recently, the Hamiltonian BRST
setting has been enriched with the topic of constructing consistent
interactions in gauge theories by means of the deformation technique and
local cohomologies \cite{39}--\cite{42}.

In this paper we investigate the consistent Hamiltonian interactions that
can be introduced between a set of vector fields and a system of matter
fields with the help of cohomological BRST arguments combined with the
deformation technique. This approach represents an extension of our former
results exposed in \cite{43} related to the abelian case. Our method goes as
follows. We begin with a ``free'' action written as the sum between the
action for a set of vector fields and an action describing a matter theory,
and construct the corresponding Hamiltonian BRST symmetry $s$, that simply
decomposes into $s=\delta +\gamma $, with $\delta $ the Koszul-Tate
differential and $\gamma $ the exterior derivative along the gauge orbits.
Its non-trivial action is essentially due to the first-class constraints
involving the vector fields. It has been shown in \cite{39}--\cite{42} that
the Hamiltonian problem of introducing consistent interactions in gauge
theories can be reformulated as a deformation problem of the BRST charge and
BRST-invariant Hamiltonian of a starting ``free'' theory. Following this
line, we firstly compute the deformed BRST charge. This necessitates the
(co)homological spaces $H\left( \gamma \right) $ and $H\left( \delta |%
\mathrm{\tilde{d}}\right) $, where $\mathrm{\tilde{d}}=\mathrm{d}%
x^{i}\partial _{i}$ represents the spatial part of the exterior space-time
derivative. Based on these (co)homologies we obtain that the deformed BRST
charge can be taken non-vanishing only at order one in the coupling
constant. The consistency of the first-order deformation requires that the
deformed first-class constraints form a Lie algebra in the Poisson (Dirac)
bracket. Secondly, we investigate the equations responsible for the
deformation of the BRST-invariant Hamiltonian. The first-order deformation
equation reveals two different types of couplings. One involves only the
vector fields and their momenta, and requires no further assumptions. The
other demands that the matter theory should display some conserved
Hamiltonian currents, equal in number to the number of vector fields.
Consequently, it follows that the second type of couplings (between vector
and matter fields) is of the form $j_{a}^{\mu }A_{\mu }^{a}$, where $%
j_{a}^{\mu }$ denote the above mentioned conserved Hamiltonian currents. The
equation that governs the second-order deformation of the BRST-invariant
Hamiltonian definitely outputs the spatial part of the quartic vertex of
pure Yang-Mills theory, and eventually other couplings involving both vector
and matter fields. The appearance of the last type of couplings depends on
the behaviour of the conserved currents under the gauge transformations
generated by the deformed first-class constraints. Thus, if the spatial part
of these currents, $j_{a}^{i}$, transform according to the adjoint
representation of the Lie gauge algebra, then there are no second-order
couplings between vector and matter fields, and, meanwhile, all types of
three- and higher-order deformations can be taken to vanish. In the opposite
case, at least the second-order deformation implying vector and matter
fields is non-trivial, but in principle there might be other relevant
higher-order interactions as well.

The paper is organized in seven sections. Section 2 briefly formulates the
analysis of consistent Hamiltonian interactions that can be added to a
``free'' theory without changing its number of gauge symmetries as a
deformation problem of the corresponding BRST charge and BRST-invariant
Hamiltonian, finally expressed in terms of the so-called main equations. In
Section 3 we determine the ``free'' Hamiltonian BRST differential. Based on
this, in Sections 4 and 5 we derive the deformed BRST charge, respectively,
the deformed BRST-invariant Hamiltonian by means of cohomological
techniques. In Section 6 we apply our procedure to two cases of interest,
where the role of the matter fields is played by a set of scalar fields,
respectively, by a collection of Dirac fields. Section 7 ends the paper with
some conclusions.

\section{Main Hamiltonian deformation equations}

We assume a ``free'' Lagrangian theory subject to some gauge
transformations. All the information on its Lagrangian gauge structure is
contained in the solution to the master equation. It has been shown that the
deformation of this solution leads to consistent interactions among fields
with gauge freedom \cite{5}. In the framework of the Hamiltonian setting,
the structure of a given gauge theory is entirely determined by two
quantities: the BRST charge and the BRST-invariant Hamiltonian.
Similar to the
Lagrangian deformation procedure, we can then reformulate the problem of
constructing consistent Hamiltonian interactions in terms of the deformation
of both the BRST charge and the BRST-invariant Hamiltonian.

As long as the interactions can be constructed in a consistent manner, the
BRST charge of a given ``free'' theory, $\Omega _{0}$, can be deformed as 
\begin{eqnarray}
&&\Omega _{0}\rightarrow \Omega =\Omega _{0}+\mathrm{g}\int \mathrm{d}%
^{D-1}x\;\omega _{1}+\mathrm{g}^{2}\int \mathrm{d}^{D-1}x\;\omega
_{2}+O\left( \mathrm{g}^{3}\right) =  \nonumber \\
&&\Omega _{0}+\mathrm{g}\Omega _{1}+\mathrm{g}^{2}\Omega _{2}+O\left( 
\mathrm{g}^{3}\right) ,  \label{s1}
\end{eqnarray}
where $\Omega $ verifies the equation 
\begin{equation}
\left[ \Omega ,\Omega \right] =0,  \label{s2}
\end{equation}
and the symbol $\left[ ,\right] $ means either the Poisson or the Dirac
bracket. (If the starting theory is purely first-class, we work with the
Poisson bracket; if second-class constraints are also present, then we
eliminate them, and use the Dirac bracket instead.) By projecting the
equation (\ref{s2}) on various powers in the deformation parameter (coupling
constant) $\mathrm{g}$, we arrive to the tower of equations 
\begin{equation}
\left[ \Omega _{0},\Omega _{0}\right] =0,  \label{s3}
\end{equation}
\begin{equation}
2\left[ \Omega _{0},\Omega _{1}\right] =0,  \label{s4}
\end{equation}
\begin{equation}
2\left[ \Omega _{0},\Omega _{2}\right] +\left[ \Omega _{1},\Omega
_{1}\right] =0,  \label{s5}
\end{equation}
\[
\vdots 
\]
Equation (\ref{s3}) is satisfied by assumption, while the resolution of the
remaining equations in terms of the ``free'' BRST differential leads to the
pieces $\left( \Omega _{k}\right) _{k>0}$. With the deformed BRST charge at
hand, we deform the BRST-invariant Hamiltonian of the ``free'' theory, $H_{0%
\mathrm{B}}$, like 
\begin{eqnarray}
&&H_{0\mathrm{B}}\rightarrow H_{\mathrm{B}}=H_{0\mathrm{B}}+\mathrm{g}\int 
\mathrm{d}^{D-1}x\;h_{1}+\mathrm{g}^{2}\int \mathrm{d}^{D-1}x\;h_{2}+O\left( 
\mathrm{g}^{3}\right) =  \nonumber \\
&&H_{0\mathrm{B}}+\mathrm{g}H_{1}+\mathrm{g}^{2}H_{2}+O\left( \mathrm{g}%
^{3}\right) ,  \label{s6}
\end{eqnarray}
and ask that it obeys the relation 
\begin{equation}
\left[ H_{\mathrm{B}},\Omega \right] =0,  \label{s7}
\end{equation}
which implements that $H_{\mathrm{B}}$ is indeed the BRST-invariant
Hamiltonian of the deformed system. Equation (\ref{s7}) can also be
investigated order by order in the deformation parameter $\mathrm{g}$,
giving 
\begin{equation}
\left[ H_{0\mathrm{B}},\Omega _{0}\right] =0,  \label{s8}
\end{equation}
\begin{equation}
\left[ H_{0\mathrm{B}},\Omega _{1}\right] +\left[ H_{1},\Omega _{0}\right]
=0,  \label{s9}
\end{equation}
\begin{equation}
\left[ H_{0\mathrm{B}},\Omega _{2}\right] +\left[ H_{1},\Omega _{1}\right]
+\left[ H_{2},\Omega _{0}\right] =0,  \label{s10}
\end{equation}
\[
\vdots 
\]
Equation (\ref{s8}) is again satisfied by hypothesis, while the others offer
the components $\left( H_{k}\right) _{k>0}$. Equations (\ref{s3}-- \ref{s5}%
), etc., and (\ref{s8}--\ref{s10}), etc., govern the Hamiltonian BRST
deformation treatment, and will be called in the sequel the main equations.

\section{Free BRST differential}

We begin with a ``free'' action written as the sum between the action for a
set of vector fields and an action describing a matter theory. We assume
that the matter fields possess no gauge invariances of their own. The
Hamiltonian canonical variables are denoted by $\left( A_{\mu }^{a},\pi
_{a}^{\mu },y^{\alpha _{0}}\right) $, where $\left( A_{\mu }^{a},\pi
_{a}^{\mu }\right) $ correspond to the vector fields, while $y^{\alpha _{0}}$
describe the matter theory. The non-vanishing fundamental Poisson (Dirac)
brackets are taken under the form 
\begin{equation}
\left[ A_{\mu }^{a},\pi _{b}^{\nu }\right] =\delta _{\mu }^{\;\;\nu }\delta
_{\;\;b}^{a},\;\left[ y^{\alpha _{0}},y^{\beta _{0}}\right] =\omega ^{\alpha
_{0}\beta _{0}},  \label{m1}
\end{equation}
with $\omega ^{\alpha _{0}\beta _{0}}$ an invertible matrix (the
distributional character was omitted for simplicity sake). Due to the
presence of the vector fields, the system is subject to the irreducible
first-class constraints 
\begin{equation}
G_{1a}\equiv \pi _{a}^{0}\approx 0,\;G_{2a}\equiv -\partial _{i}\pi
_{a}^{i}\approx 0,  \label{m2}
\end{equation}
and is endowed with the first-class Hamiltonian 
\begin{equation}
H_{0}=\int \mathrm{d}^{D-1}x\left( \frac{1}{2}\pi _{ia}\pi _{i}^{a}+\frac{1}{%
4}F_{ij}^{a}F_{a}^{ij}-A_{0}^{a}\partial _{i}\pi _{a}^{i}+\bar{h}_{0}\left(
y^{\alpha _{0}},\partial _{i}y^{\alpha _{0}}\right) \right) ,  \label{m3}
\end{equation}
where $\bar{H}_{0}=\int \mathrm{d}^{D-1}x\;\bar{h}_{0}\left( y^{\alpha
_{0}},\partial _{i}y^{\alpha _{0}}\right) $ represent the canonical
Hamiltonian of the purely matter theory. The BRST charge of this system is
given by 
\begin{equation}
\Omega _{0}=\int \mathrm{d}^{D-1}x\left( \pi _{a}^{0}\eta _{1}^{a}-\left(
\partial _{i}\pi _{a}^{i}\right) \eta _{2}^{a}\right) ,  \label{m4}
\end{equation}
the accompanying BRST-invariant Hamiltonian being 
\begin{equation}
H_{0\mathrm{B}}=H_{0}+\int \mathrm{d}^{D-1}x\;\eta _{1}^{a}\mathcal{P}_{2a}.
\label{m5}
\end{equation}
In (\ref{m4}-\ref{m5}) $\left( \eta _{1}^{a},\eta _{2}^{a}\right) $ are the
fermionic ghost number one Hamiltonian ghosts, while $\left( \mathcal{P}%
_{1a},\mathcal{P}_{2a}\right) $ stand for the associated antighosts. The
BRST complex is graded by the ghost number ($\mathrm{gh}$), defined like the
difference between the pure ghost number ($\mathrm{pgh}$) and the antighost
number ($\mathrm{antigh}$), with 
\begin{equation}
\mathrm{pgh}\left( A_{\mu }^{a}\right) =\mathrm{pgh}\left( \pi _{a}^{\mu
}\right) =\mathrm{pgh}\left( y^{\alpha _{0}}\right) =0,  \label{ma5}
\end{equation}
\begin{equation}
\mathrm{pgh}\left( \eta _{1}^{a}\right) =\mathrm{pgh}\left( \eta
_{2}^{a}\right) =1,\;\mathrm{pgh}\left( \mathcal{P}_{1a}\right) =\mathrm{pgh}%
\left( \mathcal{P}_{2a}\right) =0,  \label{ma6}
\end{equation}
\begin{equation}
\mathrm{antigh}\left( A_{\mu }^{a}\right) =\mathrm{antigh}\left( \pi
_{a}^{\mu }\right) =\mathrm{antigh}\left( y^{\alpha _{0}}\right) =0,
\label{ma7}
\end{equation}
\begin{equation}
\mathrm{antigh}\left( \eta _{1}^{a}\right) =\mathrm{antigh}\left( \eta
_{2}^{a}\right) =0,\;\mathrm{antigh}\left( \mathcal{P}_{1a}\right) =\mathrm{%
antigh}\left( \mathcal{P}_{2a}\right) =1.  \label{ma8}
\end{equation}
The ``free'' BRST symmetry $s\bullet =\left[ \bullet ,\Omega _{0}\right] $
splits as 
\begin{equation}
s=\delta +\gamma ,  \label{m6}
\end{equation}
where $\delta $ is the Koszul-Tate differential, graded according to the
antighost number ($\mathrm{antigh}\left( \delta \right) =-1$, $\mathrm{antigh%
}\left( \gamma \right) =0$), and $\gamma $ is the exterior longitudinal
derivative along the gauge orbits, graded in terms of the pure ghost number (%
$\mathrm{pgh}\left( \gamma \right) =1$, $\mathrm{pgh}\left( \delta \right) =0
$). These operators act on the variables from the BRST complex via the
definitions 
\begin{equation}
\delta A_{\mu }^{a}=0,\;\delta \pi _{a}^{\mu }=0,\;\delta y^{\alpha _{0}}=0,
\label{m7}
\end{equation}
\begin{equation}
\delta \eta _{1}^{a}=\delta \eta _{2}^{a}=0,\;\delta \mathcal{P}_{1a}=-\pi
_{a}^{0},\;\delta \mathcal{P}_{2a}=\partial _{i}\pi _{a}^{i},  \label{m8}
\end{equation}
\begin{equation}
\gamma A_{0}^{a}=\eta _{1}^{a},\;\gamma A_{i}^{a}=\partial _{i}\eta
_{2}^{a},\;\gamma \pi _{a}^{\mu }=0,\;\gamma y^{\alpha _{0}}=0,  \label{m9}
\end{equation}
\begin{equation}
\gamma \eta _{1}^{a}=\gamma \eta _{2}^{a}=0,\;\gamma \mathcal{P}_{1a}=\gamma 
\mathcal{P}_{2a}=0,  \label{m10}
\end{equation}
that will be used in the sequel during the deformation process.

\section{Deformation of the BRST charge}

In this section we analyse the main equations (\ref{s4}--\ref{s5}), etc.,
that describe the deformation of the ``free'' BRST charge. The equation (\ref
{s4}) written in a local form becomes 
\begin{equation}
s\omega _{1}=\partial _{i}k^{i},  \label{m11}
\end{equation}
for some local $k^{i}$. In order to solve the equation (\ref{m11}), we
develop $\omega _{1}$ according to the antighost number 
\begin{equation}
\omega _{1}=\stackrel{\left( 0\right) }{\omega }_{1}+\stackrel{\left(
1\right) }{\omega }_{1}+\cdots +\stackrel{\left( J\right) }{\omega }_{1},
\label{m12}
\end{equation}
with 
\begin{equation}
\mathrm{antigh}\left( \stackrel{\left( I\right) }{\omega }_{1}\right) =I,\;%
\mathrm{pgh}\left( \stackrel{\left( I\right) }{\omega }_{1}\right) =1,
\label{m13}
\end{equation}
where the last term in (\ref{m12}) can be assumed to be annihilated by $%
\gamma $, $\gamma \stackrel{\left( J\right) }{\omega }_{1}=0$. Thus, we need
to know the cohomology of $\gamma $, $H\left( \gamma \right) $, in order to
output $\stackrel{\left( J\right) }{\omega }_{1}$. Looking at the
definitions (\ref{m9}--\ref{m10}), it results that $H\left( \gamma \right) $
is generated by $F_{ij}^{a}$, $\pi _{a}^{\mu }$, $y^{\alpha _{0}}$, $%
\mathcal{P}_{1a}$, $\mathcal{P}_{2a}$ and their spatial derivatives, as well
as by the undifferentiated ghosts $\eta _{2}^{a}$. The ghosts $\eta _{1}^{a}$
do not enter the cohomology of $\gamma $ as they are $\gamma $-exact by
virtue of the former definitions in (\ref{m9}). As a consequence, the
general solution to the equation $\gamma \alpha =0$, can be represented (up
to a trivial term) as 
\begin{equation}
\alpha =\alpha _{M}\left( \left[ F_{ij}^{a}\right] ,\left[ \pi _{a}^{\mu
}\right] ,\left[ y^{\alpha _{0}}\right] ,\left[ \mathcal{P}_{1a}\right]
,\left[ \mathcal{P}_{2a}\right] \right) e^{M}\left( \eta _{2}^{a}\right) ,
\label{m14}
\end{equation}
with $e^{M}\left( \eta _{2}^{a}\right) $ a basis in the finite-dimensional
space of polynomials in the ghosts $\eta _{2}^{a}$, while the notation $%
f\left[ q\right] $ signifies that $f$ depends on $q$ and its derivatives up
to a finite order. Then, the equation $\gamma \stackrel{\left( J\right) }{%
\omega }_{1}=0$ possesses the solution 
\begin{equation}
\stackrel{\left( J\right) }{\omega }_{1}=\tilde{\omega}_{J}\left( \left[
F_{ij}^{a}\right] ,\left[ \pi _{a}^{\mu }\right] ,\left[ y^{\alpha
_{0}}\right] ,\left[ \mathcal{P}_{1a}\right] ,\left[ \mathcal{P}_{2a}\right]
\right) e^{J+1}\left( \eta _{2}^{a}\right) ,  \label{m15}
\end{equation}
where $\mathrm{pgh}\left( e^{J+1}\left( \eta _{2}^{a}\right) \right) =J+1$
and $\mathrm{antigh}\left( \tilde{\omega}_{J}\right) =J$. Related to the
component of antighost number $\left( J-1\right) $, the equation (\ref{m11})
becomes 
\begin{equation}
\delta \stackrel{\left( J\right) }{\omega }_{1}+\gamma \stackrel{\left(
J-1\right) }{\omega }_{1}=\partial _{i}m^{i}.  \label{m16}
\end{equation}
For the last equation to display solutions it is necessary that $\tilde{%
\omega}_{J}$ belongs to $H_{J}\left( \delta |\mathrm{\tilde{d}}\right) $.
However, using the general results from \cite{16}--\cite{16a} adapted to our
situation, we have that 
\begin{equation}
H_{J}\left( \delta |\mathrm{\tilde{d}}\right) =0,\;\mathrm{for}\;J>1.
\label{m17}
\end{equation}
This implies that we can assume that the development (\ref{m12}) stops after
the first two terms 
\begin{equation}
\omega _{1}=\stackrel{\left( 0\right) }{\omega }_{1}+\stackrel{\left(
1\right) }{\omega }_{1},  \label{m18}
\end{equation}
with $\gamma \stackrel{\left( 1\right) }{\omega }_{1}=0$. On behalf of (\ref
{m15}), we find that $\stackrel{\left( 1\right) }{\omega }_{1}=\tilde{\omega}%
_{ab}\eta _{2}^{a}\eta _{2}^{b}$, where the coefficients $\tilde{\omega}%
_{ab}=-\tilde{\omega}_{ba}$ pertain to $H_{1}\left( \delta |\mathrm{\tilde{d}%
}\right) $, i.e., 
\begin{equation}
\delta \tilde{\omega}_{ab}=\partial _{i}n_{ab}^{i}.  \label{m19}
\end{equation}
On the other hand, the general representative of $H_{1}\left( \delta |%
\mathrm{\tilde{d}}\right) $ is $\lambda =\lambda ^{a}\mathcal{P}_{2a}$ (see (%
\ref{m8})), with constant $\lambda ^{a}$. Then, we can write that $\tilde{%
\omega}_{ab}=\frac{1}{2}f_{\;\;ab}^{c}\mathcal{P}_{2c}$, where $%
f_{\;\;ab}^{c}=-f_{\;\;ba}^{c}$ are constant, hence 
\begin{equation}
\stackrel{\left( 1\right) }{\omega }_{1}=\frac{1}{2}f_{\;\;ab}^{c}\mathcal{P}%
_{2c}\eta _{2}^{a}\eta _{2}^{b}.  \label{m20}
\end{equation}
It follows that the solution to the equation associated with the antighost
number equal to zero, $\delta \stackrel{\left( 1\right) }{\omega }%
_{1}+\gamma \stackrel{\left( 0\right) }{\omega }_{1}=\partial _{i}q^{i}$,
reads as 
\begin{equation}
\stackrel{\left( 0\right) }{\omega }_{1}=\left( f_{\;\;ab}^{c}\pi
_{c}^{i}A_{i}^{b}+b_{a}\left( y^{\alpha _{0}},\partial _{i}y^{\alpha
_{0}}\right) \right) \eta _{2}^{a},  \label{m21}
\end{equation}
where the bosonic functions $b_{a}\left( y^{\alpha _{0}},\partial
_{i}y^{\alpha _{0}}\right) $ are arbitrary at this stage. Combining the
above results, we obtain the first-order deformation of the BRST charge
under the form 
\begin{equation}
\Omega _{1}=\int \mathrm{d}^{D-1}x\left( \left( f_{\;\;ab}^{c}\pi
_{c}^{i}A_{i}^{b}+b_{a}\left( y^{\alpha _{0}},\partial _{i}y^{\alpha
_{0}}\right) \right) \eta _{2}^{a}+\frac{1}{2}f_{\;\;ab}^{c}\mathcal{P}%
_{2c}\eta _{2}^{a}\eta _{2}^{b}\right) .  \label{m22}
\end{equation}

Next, we investigate its second-order deformation. By direct computation we
get 
\begin{eqnarray}
&&\left[ \Omega _{1},\Omega _{1}\right] =f_{\;\;\left[ ab\right.
}^{d}f_{\;\;\left. c\right] d}^{e}\int \mathrm{d}^{D-1}x\left( \eta
_{2}^{a}\eta _{2}^{b}\pi _{e}^{i}A_{i}^{c}-\frac{1}{3}\eta _{2}^{a}\eta
_{2}^{b}\eta _{2}^{c}\mathcal{P}_{2e}\right) +  \nonumber \\
&&\int \mathrm{d}^{D-1}x\;\mathrm{d}^{D-1}x^{\prime }\left( \left[
b_{a}\left( x\right) ,b_{b}\left( x^{\prime }\right) \right] -\right.  
\nonumber \\
&&\left. f_{\;\;ab}^{c}b_{c}\left( x\right) \delta ^{D-1}\left( x-x^{\prime
}\right) \right) \eta _{2}^{a}\left( x\right) \eta _{2}^{b}\left( x^{\prime
}\right) .  \label{m23}
\end{eqnarray}
Equation (\ref{s5}) asks that $\left[ \Omega _{1},\Omega _{1}\right] $ is $s$%
-exact. However, from (\ref{m23}) we observe that this requirement cannot be
fulfilled, so $\left[ \Omega _{1},\Omega _{1}\right] $ should vanish. This
holds if and only if the following conditions are simultaneously satisfied
\begin{equation}
f_{\;\;\left[ ab\right. }^{d}f_{\;\;\left. c\right] d}^{e}=0,  \label{m24}
\end{equation}
\begin{equation}
\left[ b_{a}\left( x\right) ,b_{b}\left( x^{\prime }\right) \right]
=f_{\;\;ab}^{c}b_{c}\left( x\right) \delta ^{D-1}\left( x-x^{\prime }\right)
.  \label{m25}
\end{equation}
The formula (\ref{m24}) shows that the antisymmetric constants $%
f_{\;\;ab}^{c}$ satisfy Jacobi's identity, being thus the structure
constants of a Lie algebra. Formula (\ref{m25}) restricts the form of the
functions $b_{a}$ in the sense that they form a Lie algebra in the Poisson
(Dirac) bracket, with structure constants $f_{\;\;ab}^{c}$. Due to the fact
that $\left[ \Omega _{1},\Omega _{1}\right] =0$, we deduce that we can take $%
\Omega _{2}=\Omega _{3}=\cdots =0$. Now, we solve the equations (\ref{m25}).
They possess solutions if and only if the fields $y^{\alpha _{0}}$ split
into two subsets 
\begin{equation}
y^{\alpha _{0}}=\left( y^{\alpha _{1}},z_{\alpha _{1}}\right) ,  \label{m26}
\end{equation}
with the properties 
\begin{equation}
\left[ y^{\alpha _{1}},y^{\beta _{1}}\right] =0,\;\left[ z_{\alpha
_{1}},z_{\beta _{1}}\right] =0,\;\left[ y^{\alpha _{1}},z_{\beta
_{1}}\right] =\sigma _{\;\;\beta _{1}}^{\alpha _{1}},  \label{m27}
\end{equation}
for some invertible matrices $\sigma _{\;\;\beta _{1}}^{\alpha _{1}}$ (the
distributional character has been again omitted). Under these circumstances,
we find that 
\begin{equation}
b_{a}=z_{\alpha _{1}}T_{a\;\;\beta _{1}}^{\alpha _{1}}\bar{\sigma}_{\;\;\rho
_{1}}^{\beta _{1}}y^{\rho _{1}},  \label{m28}
\end{equation}
with $\bar{\sigma}_{\;\;\rho _{1}}^{\beta _{1}}$ the inverse of $\sigma
_{\;\;\beta _{1}}^{\alpha _{1}}$, and $T_{a\;\;\beta _{1}}^{\alpha _{1}}$
some constant matrices, subject to the conditions 
\begin{equation}
T_{a\;\;\beta _{1}}^{\alpha _{1}}T_{b\;\;\rho _{1}}^{\beta
_{1}}-T_{b\;\;\beta _{1}}^{\alpha _{1}}T_{a\;\;\rho _{1}}^{\beta
_{1}}=f_{\;\;ab}^{c}T_{c\;\;\rho _{1}}^{\alpha _{1}}.  \label{m29}
\end{equation}
The presence of $\bar{\sigma}_{\;\;\rho _{1}}^{\beta _{1}}$ in (\ref{m28})
may in principle lead to the loss of locality. As we restrict ourselves to
local deformations only, we consider the case of constant $\sigma
_{\;\;\beta _{1}}^{\alpha _{1}}$. Therefore, the deformed BRST charge
consistent to all orders in the deformation parameter is expressed by 
\begin{eqnarray}
&&\Omega =\int \mathrm{d}^{D-1}x\left( \pi _{a}^{0}\eta _{1}^{a}-\left(
\left( \mathrm{D}_{i}\right) _{a}^{\;\;b}\pi _{b}^{i}-\mathrm{g}z_{\alpha
_{1}}T_{a\;\;\beta _{1}}^{\alpha _{1}}\bar{\sigma}_{\;\;\rho _{1}}^{\beta
_{1}}y^{\rho _{1}}\right) \eta _{2}^{a}+\right.   \nonumber \\
&&\left. \frac{1}{2}\mathrm{g}f_{\;\;ab}^{c}\mathcal{P}_{2c}\eta
_{2}^{a}\eta _{2}^{b}\right) ,  \label{m30}
\end{eqnarray}
with $\left( \mathrm{D}_{i}\right) _{a}^{\;\;b}=\delta _{a}^{\;\;b}\partial
_{i}-\mathrm{g}f_{\;\;ac}^{b}A_{i}^{c}$. From $\Omega $ we can gather
information on the deformed constraints and modified gauge algebra. Indeed,
the term in $\Omega $ linear in the ghosts $\eta _{2}^{a}$ gives rise to the
deformed secondary constraints 
\begin{equation}
\gamma _{2a}\equiv -\left( \mathrm{D}_{i}\right) _{a}^{\;\;b}\pi _{b}^{i}+%
\mathrm{g}z_{\alpha _{1}}T_{a\;\;\beta _{1}}^{\alpha _{1}}\bar{\sigma}%
_{\;\;\rho _{1}}^{\beta _{1}}y^{\rho _{1}}\approx 0,  \label{m31}
\end{equation}
while the term linear in the antighosts shows that these constraint
functions form a Lie algebra in the Poisson (Dirac) bracket 
\begin{equation}
\left[ \gamma _{2a},\gamma _{2b}\right] =\mathrm{g}f_{\;\;ab}^{c}\gamma
_{2c},  \label{m32}
\end{equation}
with the structure constants $f_{\;\;ab}^{c}$. This completes the
deformation procedure of the BRST charge for a collection of vector fields
and matter fields.

\section{Deformation of the BRST-invariant Hamiltonian}

Further, we approach the equations (\ref{s9}--\ref{s10}), etc., that control
the deformation of the BRST-invariant Hamiltonian. By direct computation we
find that the first term in the left hand-side of (\ref{s9}) reads as 
\begin{eqnarray}
&&\left[ H_{0\mathrm{B}},\Omega _{1}\right] =-s\int \mathrm{d}^{D-1}x\left(
f_{\;\;bc}^{a}\left( A_{0}^{b}\left( A_{i}^{c}\pi _{a}^{i}+\eta _{2}^{c}%
\mathcal{P}_{2a}\right) -\frac{1}{2}A_{i}^{b}A_{j}^{c}F_{a}^{ij}\right)
+b_{a}A_{0}^{a}\right) +  \nonumber \\
&&\int \mathrm{d}^{D-1}x\;\left[ \bar{H}_{0},b_{a}\right] \eta _{2}^{a}.
\label{m33}
\end{eqnarray}
We notice that the last term in the right hand-side of (\ref{m33}) is
clearly not $s$-exact, so it must be compensated by a corresponding term
from $\left[ H_{1},\Omega _{0}\right] $, which can be accomplished if we
take $H_{1}$ of the form 
\begin{equation}
H_{1}=\int \mathrm{d}^{D-1}x\left( f_{\;\;bc}^{a}\left( A_{0}^{b}\left(
A_{i}^{c}\pi _{a}^{i}+\eta _{2}^{c}\mathcal{P}_{2a}\right) -\frac{1}{2}%
A_{i}^{b}A_{j}^{c}F_{a}^{ij}\right) +b_{a}A_{0}^{a}+j\right) .  \label{m34}
\end{equation}
The function $j$ involves the vector fields $A_{i}^{a}$ and the matter
fields, and, moreover, we ask that it fulfills the equation $\int \mathrm{d}%
^{D-1}x\left( \left[ \bar{H}_{0},b_{a}\right] \eta _{2}^{a}+\left[ j,\Omega
_{0}\right] \right) =0$, or, equivalently 
\begin{equation}
\left[ \bar{H}_{0}\left( x_{0}\right) ,b_{a}\left( x\right) \right] \eta
_{2}^{a}\left( x\right) +\left[ j\left( x\right) ,\Omega _{0}\left(
x_{0}\right) \right] =\partial _{i}k^{i}\left( x\right) .  \label{m35}
\end{equation}
Using (\ref{m4}), the last equation becomes 
\begin{equation}
\left[ \bar{H}_{0}\left( x_{0}\right) ,b_{a}\left( x\right) \right] \eta
_{2}^{a}\left( x\right) +\int \mathrm{d}^{D-1}x^{\prime }\;\left[ j\left(
x\right) ,\pi _{a}^{i}\left( x^{\prime }\right) \right] \partial _{i}\eta
_{2}^{a}\left( x^{\prime }\right) =\partial _{i}k^{i}\left( x\right) .
\label{m36}
\end{equation}
In order to restrain the left hand-side to a total derivative, it is
necessary that the function $j\left( x\right) $ is linear in the fields $%
A_{i}^{a}$ because the term $\left[ \bar{H}_{0}\left( x_{0}\right)
,b_{a}\left( x\right) \right] $ does not involve the vector fields. Thus, we
can write 
\begin{equation}
j=j_{a}^{i}A_{i}^{a},  \label{m37}
\end{equation}
where $j_{a}^{i}$ depends only on the matter fields and their derivatives.
Consequently, the equation (\ref{m36}) takes the form 
\begin{equation}
\left[ \bar{H}_{0}\left( x_{0}\right) ,b_{a}\left( x\right) \right] \eta
_{2}^{a}\left( x\right) +j_{a}^{i}\left( x\right) \partial _{i}\eta
_{2}^{a}\left( x\right) =\partial _{i}k^{i}\left( x\right) .  \label{m38}
\end{equation}
The left hand-side of (\ref{m38}) reduces to a total derivative if and only
if 
\begin{equation}
\left[ \bar{H}_{0}\left( x_{0}\right) ,b_{a}\left( x\right) \right]
=\partial _{i}j_{a}^{i}\left( x\right) .  \label{m39}
\end{equation}
By means of the Hamilton equations with respect to the matter fields ($%
\partial _{0}F\left( x\right) =\left[ F\left( x\right) ,\bar{H}_{0}\left(
x_{0}\right) \right] $), from (\ref{m39}) we derive that 
\begin{equation}
\partial _{0}b_{a}\left( x\right) +\partial _{i}j_{a}^{i}\left( x\right) =0,
\label{m40}
\end{equation}
which expresses nothing but the conservation of the Hamiltonian currents%
\footnote{%
By Hamiltonian currents we understand a set of functions $j_{a}^{\mu
}=\left( b_{a},j_{a}^{i}\right) $ that satisfy the equation (\ref{m40}) when
the Hamiltonian equations of motion hold. In general, these currents do not
display a manifestly covariant form. However, if we express these currents
only in terms of the fields (via the elimination of the momenta on their
equations of motion), we infer precisely their Lagrangian form, which is
clearly covariant.} $j_{a}^{\mu }=\left( b_{a},j_{a}^{i}\right) $. In
conclusion, the consistency of the first-order deformation of the
BRST-invariant Hamiltonian requires that the matter theory displays some
conserved currents, equal in number with the number of vector fields $A_{\mu
}^{a}$. In the following we assume that this request is fulfilled. Then, the
first-order deformation of the BRST-invariant Hamiltonian is given by 
\begin{equation}
H_{1}=\int \mathrm{d}^{D-1}x\left( f_{\;\;bc}^{a}\left( A_{0}^{b}\left(
A_{i}^{c}\pi _{a}^{i}+\eta _{2}^{c}\mathcal{P}_{2a}\right) -\frac{1}{2}%
A_{i}^{b}A_{j}^{c}F_{a}^{ij}\right) +j_{a}^{\mu }A_{\mu }^{a}\right) ,
\label{m41}
\end{equation}
where $j_{a}^{\mu }$ stand for the conserved Hamiltonian currents mentioned
in the above.

Now, we pass at the second-order deformation, described by the equation (\ref
{s10}). Making use of the formulas (\ref{m30}) and (\ref{m41}), it results
that 
\begin{eqnarray}
&&\left[ H_{1},\Omega _{1}\right] =s\left( \int \mathrm{d}^{D-1}x\;\frac{1}{4%
}f_{\;\;bc}^{a}f_{\;\;de}^{c}A^{ib}A_{a}^{j}A_{i}^{d}A_{j}^{e}\right) + 
\nonumber \\
&&\int \mathrm{d}^{D-1}x\left( \left[ j_{b}^{i},b_{a}\right]
+f_{\;\;ab}^{c}j_{c}^{i}\right) A_{i}^{b}\eta _{2}^{a}.  \label{m42}
\end{eqnarray}
Looking at the form of (\ref{m42}), we remark that two important cases
appear.

a) If the currents $j_{a}^{i}$ transform under the deformed gauge
transformations (generated by the deformed first-class constraints)\footnote{%
As $j_{b}^{i}$ depend only on the matter fields and their derivatives, we
find that their deformed gauge transformations are indeed $\bar{\delta}%
_{\epsilon }j_{b}^{i}=\left[ j_{b}^{i},\gamma _{2a}\right] \epsilon
_{2}^{a}=g\left[ j_{b}^{i},b_{a}\right] \epsilon _{2}^{a}$.} according to
the adjoint representation of the Lie gauge algebra 
\begin{equation}
\left[ j_{b}^{i},b_{a}\right] +f_{\;\;ab}^{c}j_{c}^{i}=0,  \label{m43}
\end{equation}
then the second-order deformation of the BRST-invariant Hamiltonian will be 
\begin{equation}
H_{2}=-\frac{1}{4}\int \mathrm{d}^{D-1}x\;f_{\;\;bc}^{a}f_{\;%
\;de}^{c}A^{ib}A_{a}^{j}A_{i}^{d}A_{j}^{e}.  \label{m44}
\end{equation}
As $\Omega _{2}=0$ and $\left[ H_{2},\Omega _{1}\right] =0$, the third-order
deformation equation is satisfied with the choice $H_{3}=0$. The equations
for higher-order deformations will then be checked for 
\begin{equation}
H_{4}=H_{5}=\cdots =0.  \label{m45}
\end{equation}

b) In the opposite situation 
\begin{equation}
\left[ j_{b}^{i},b_{a}\right] +f_{\;\;ab}^{c}j_{c}^{i}\neq 0,  \label{m46}
\end{equation}
there appear non-trivial higher-order deformations that imply interactions
among vector fields and matter fields.

In both cases, the deformed BRST-invariant Hamiltonian has the general form 
\begin{eqnarray}
&&H_{\mathrm{B}}=\int \mathrm{d}^{D-1}x\left( \frac{1}{2}\pi _{ia}\pi
_{i}^{a}+\frac{1}{4}\tilde{F}_{ij}^{a}\tilde{F}_{a}^{ij}-A_{0}^{a}\left( 
\mathrm{D}_{i}\right) _{a}^{\;\;b}\pi _{b}^{i}+\bar{h}_{0}\left( y^{\alpha
_{0}},\partial _{i}y^{\alpha _{0}}\right) +\right.   \nonumber \\
&&\left. \mathrm{g}j_{a}^{\mu }A_{\mu }^{a}+\left( \eta _{1}^{a}-\mathrm{g}%
f_{\;\;bc}^{a}\eta _{2}^{b}A_{0}^{c}\right) \mathcal{P}_{2a}\right) +O\left( 
\mathrm{g}^{2}\right) ,  \label{m47}
\end{eqnarray}
where 
\begin{equation}
\tilde{F}_{ij}^{a}=F_{ij}^{a}-\mathrm{g}f_{\;\;bc}^{a}A_{i}^{b}A_{j}^{c},
\label{m48}
\end{equation}
and $O\left( \mathrm{g}^{2}\right) $ is due only to the supplementary terms
present in case b).

The deformation treatment developed so far can be synthesized in three
general results as follows. First, the interaction terms involving only the
vector fields generate the Hamiltonian version of Yang-Mills theory, and the
first-order couplings between the vector fields and matter fields is of the
type $j_{a}^{\mu }A_{\mu }^{a}$, where $j_{a}^{\mu }=\left(
b_{a},j_{a}^{i}\right) $ are the conserved Hamiltonian currents
corresponding to the matter fields. Second, the secondary first-class
constraints are deformed with respect to the initial ones, and, as a
consequence, the corresponding gauge transformations will be modified.
Third, the deformed gauge algebra of first-class constraints is a Lie
algebra. Finally, a word of caution. Once the deformations related to a
given matter theory are computed, special attention should be paid to the
elimination of non-locality, as well as of triviality of the resulting
deformations. This completes our general procedure.

\section{Applications}

In the sequel we apply the general deformation procedure investigated so far
to two cases of interest, where the matter theory involves scalar fields,
respectively, Dirac fields.

\subsection{Vector fields coupled to scalar fields}

First, we analyse the consistent interactions that can be introduced among a
set of real scalar fields and a collection of vector fields. In this case
the ``free'' Lagrangian action is given by 
\begin{equation}
\tilde{S}_{0}^{\mathrm{L}}\left[ A_{\mu }^{a},\varphi ^{A}\right] =\int 
\mathrm{d}^{4}x\left( -\frac{1}{4}F_{\mu \nu }^{a}F_{a}^{\mu \nu }+\frac{1}{2%
}K_{AB}\left( \partial _{\mu }\varphi ^{A}\right) \left( \partial ^{\mu
}\varphi ^{B}\right) -V\left( \varphi ^{A}\right) \right) ,  \label{m49}
\end{equation}
where $K_{AB}$ is an invertible symmetric constant matrix. Action (\ref{m49}%
) is invariant under the gauge transformations 
\begin{equation}
\delta _{\epsilon }A_{\mu }^{a}=\partial _{\mu }\epsilon ^{a},\;\delta
_{\epsilon }\varphi ^{A}=0,  \label{m49a}
\end{equation}
and the first-class Hamiltonian is expressed by (\ref{m3}), with 
\begin{equation}
\bar{h}_{0}=\frac{1}{2}K^{AB}\Pi _{A}\Pi _{B}-\frac{1}{2}K_{AB}\left(
\partial _{j}\varphi ^{A}\right) \left( \partial ^{j}\varphi ^{B}\right)
+V\left( \varphi ^{A}\right) ,  \label{m50}
\end{equation}
the matrix $K^{AB}$ denoting the inverse of $K_{AB}$, and $\Pi _{A}$ meaning
the canonical momenta conjugated to $\varphi ^{A}$. In this situation we
have that 
\begin{equation}
y^{\alpha _{0}}=\left( \varphi ^{A},\Pi _{A}\right) ,  \label{m51}
\end{equation}
with 
\begin{equation}
\left[ \varphi ^{A},\varphi ^{B}\right] =0,\;\left[ \Pi _{A},\Pi _{B}\right]
=0,\;\left[ \varphi ^{A},\Pi _{B}\right] =\delta _{\;\;B}^{A}.  \label{m52}
\end{equation}
By performing the identifications 
\begin{equation}
y^{\alpha _{1}}\longleftrightarrow \varphi ^{A},\;z_{\alpha
_{1}}\longleftrightarrow \Pi _{A},\;\sigma _{\;\;\beta _{1}}^{\alpha
_{1}}\longleftrightarrow \delta _{\;\;B}^{A},  \label{m53}
\end{equation}
from (\ref{m28}) we obtain that the conserved charge $b_{a}$ is precisely
given by 
\begin{equation}
b_{a}=\Pi _{A}T_{a\;\;B}^{A}\varphi ^{B}.  \label{m54}
\end{equation}
With $b_{a}$ at hand, we then deduce the form of the currents $j_{a}^{i}$ by
employing the formula (\ref{m39}). On behalf of (\ref{m50}), we get that 
\begin{equation}
\left[ b_{a},\bar{H}_{0}\right] =T_{a\;\;B}^{A}K^{BC}\Pi _{A}\Pi _{C}-\frac{%
\partial V}{\partial \varphi ^{A}}T_{a\;\;B}^{A}\varphi
^{B}-K_{AC}T_{a\;\;B}^{A}\left( \partial _{i}\partial ^{i}\varphi
^{C}\right) \varphi ^{B}.  \label{m55}
\end{equation}
In order to reveal some conserved Hamiltonian currents in the matter sector,
it is necessary that 
\begin{equation}
\frac{\partial V}{\partial \varphi ^{A}}T_{a\;\;B}^{A}\varphi ^{B}=0,
\label{m56}
\end{equation}
\begin{equation}
\tilde{T}_{aBC}=-\tilde{T}_{aCB},\;\bar{T}_{a}^{AC}=-\bar{T}_{a}^{CA},
\label{m57}
\end{equation}
where $\tilde{T}_{aBC}\equiv K_{AB}T_{a\;\;C}^{A}$, $\bar{T}_{a}^{AC}\equiv
K^{BC}T_{a\;\;B}^{A}$. Inserting (\ref{m56}-\ref{m57}) in (\ref{m55}), we
arrive at 
\begin{equation}
\left[ b_{a},\bar{H}_{0}\right] +\partial _{i}\left(
K_{AC}T_{a\;\;B}^{A}\left( \partial ^{i}\varphi ^{C}\right) \varphi
^{B}\right) =0,  \label{m58}
\end{equation}
and hence the conserved currents are 
\begin{equation}
j_{a}^{i}=K_{AC}T_{a\;\;B}^{A}\left( \partial ^{i}\varphi ^{C}\right)
\varphi ^{B}.  \label{m59}
\end{equation}
Once we determined $b_{a}$ and $j_{a}^{i}$, the deformed BRST charge and the
first-order deformation of the BRST-invariant Hamiltonian are completely
constructed. Regarding the second-order deformation of the Hamiltonian, by
direct computation we deduce 
\begin{eqnarray}
&&\int \mathrm{d}^{3}x\;\left[ j_{b}^{i},b_{a}\right] A_{i}^{b}\eta
_{2}^{a}=-\int \mathrm{d}^{3}x\;f_{\;\;ab}^{c}j_{c}^{i}A_{i}^{b}\eta
_{2}^{a}+  \nonumber \\
&&s\left( \int \mathrm{d}^{3}x\;\frac{1}{2}K_{AC}T_{b\;\;B}^{A}T_{a\;%
\;E}^{C}\varphi ^{B}\varphi ^{E}A^{ai}A_{i}^{b}\right) ,  \label{m60}
\end{eqnarray}
so there are met the conditions of case b) (see (\ref{m46})). Then, we infer
that 
\begin{equation}
H_{2}=-\int \mathrm{d}^{3}x\left( \frac{1}{4}f_{\;\;bc}^{a}f_{\;%
\;de}^{c}A^{ib}A_{a}^{j}A_{i}^{d}A_{j}^{e}+\frac{1}{2}K_{AC}T_{b\;%
\;B}^{A}T_{a\;\;E}^{C}\varphi ^{B}\varphi ^{E}A^{ai}A_{i}^{b}\right) .
\label{m61}
\end{equation}
As $\left[ H_{2},\Omega _{1}\right] =0$, it follows that the third-order
deformation equation is verified for $H_{3}=0$, and similarly we can take $%
H_{4}=H_{5}=\cdots =0$.

Gathering the results derived so far, it follows that both the deformed BRST
charge and BRST-invariant Hamiltonian can respectively be written in the
form 
\begin{eqnarray}
&&\Omega =\int \mathrm{d}^{3}x\left( \pi _{a}^{0}\eta _{1}^{a}-\left( \left( 
\mathrm{D}_{i}\right) _{a}^{\;\;b}\pi _{b}^{i}-\mathrm{g}\Pi
_{A}T_{a\;\;B}^{A}\varphi ^{B}\right) \eta _{2}^{a}+\right.   \nonumber \\
&&\left. \frac{1}{2}\mathrm{g}f_{\;\;ab}^{c}\mathcal{P}_{2c}\eta
_{2}^{a}\eta _{2}^{b}\right) ,  \label{m62}
\end{eqnarray}
\begin{eqnarray}
&&H_{\mathrm{B}}=\int \mathrm{d}^{3}x\left( \frac{1}{2}\pi _{ia}\pi _{i}^{a}+%
\frac{1}{4}\tilde{F}_{ij}^{a}\tilde{F}_{a}^{ij}-A_{0}^{a}\left( \left( 
\mathrm{D}_{i}\right) _{a}^{\;\;b}\pi _{b}^{i}-\mathrm{g}\Pi
_{A}T_{a\;\;B}^{A}\varphi ^{B}\right) +\right.   \nonumber \\
&&\frac{1}{2}K^{AB}\Pi _{A}\Pi _{B}-\frac{1}{2}K_{AB}\left( \mathrm{D}%
_{j\;\;C}^{A}\varphi ^{C}\right) \left( \mathrm{D}_{\;\;D}^{jB}\varphi
^{D}\right) +V\left( \varphi ^{A}\right) +  \nonumber \\
&&\left. \left( \eta _{1}^{a}-\mathrm{g}f_{\;\;bc}^{a}\eta
_{2}^{b}A_{0}^{c}\right) \mathcal{P}_{2a}\right) ,  \label{m63}
\end{eqnarray}
where 
\begin{equation}
\mathrm{D}_{j\;\;C}^{A}=\delta _{\;\;C}^{A}\partial _{j}+\mathrm{g}%
T_{a\;\;C}^{A}A_{j}^{a}.  \label{m64}
\end{equation}
Now, we have enough information to identify the resulting interacting
theory. Only the secondary constraints are deformed 
\begin{equation}
\gamma _{2a}\equiv -\left( \left( \mathrm{D}_{i}\right) _{a}^{\;\;b}\pi
_{b}^{i}-\mathrm{g}\Pi _{A}T_{a\;\;B}^{A}\varphi ^{B}\right) \approx 0,
\label{m65}
\end{equation}
and they form a Lie algebra in the Poisson bracket, like in (\ref{m32}). The
antighost number zero piece in (\ref{m63}) 
\begin{eqnarray}
&&\tilde{H}_{0}=\int \mathrm{d}^{3}x\left( \frac{1}{2}\pi _{ia}\pi _{i}^{a}+%
\frac{1}{4}\tilde{F}_{ij}^{a}\tilde{F}_{a}^{ij}-A_{0}^{a}\left( \left( 
\mathrm{D}_{i}\right) _{a}^{\;\;b}\pi _{b}^{i}-\mathrm{g}\Pi
_{A}T_{a\;\;B}^{A}\varphi ^{B}\right) +\right.   \nonumber \\
&&\left. \frac{1}{2}K^{AB}\Pi _{A}\Pi _{B}-\frac{1}{2}K_{AB}\left( \mathrm{D}%
_{j\;\;C}^{A}\varphi ^{C}\right) \left( \mathrm{D}_{\;\;D}^{jB}\varphi
^{D}\right) +V\left( \varphi ^{A}\right) \right) ,  \label{m66}
\end{eqnarray}
represents the first-class Hamiltonian of the coupled theory. The component
from (\ref{m63}) linear in the antighosts underlines that the Poisson
brackets between the constraints and the first-class Hamiltonian are
modified like 
\begin{equation}
\left[ \tilde{H}_{0},G_{1a}\right] =\gamma _{2a},\;\left[ \tilde{H}%
_{0},\gamma _{2a}\right] =-\mathrm{g}f_{\;\;ab}^{c}A_{0}^{b}\gamma _{2c}.
\label{m67}
\end{equation}
In this way, we observe that we have obtained nothing but the Hamiltonian
version of the theory that describes the interaction between Yang-Mills
fields and a set of scalar fields.

The Lagrangian version of the interacting model can be derived in the usual
manner, yielding thus the action 
\begin{equation}
S^{\mathrm{L}}\left[ A_{\mu }^{a},\varphi ^{A}\right] =\int \mathrm{d}%
^{4}x\left( -\frac{1}{4}\tilde{F}_{\mu \nu }^{a}\tilde{F}_{a}^{\mu \nu }+%
\frac{1}{2}K_{AB}\left( \mathrm{D}_{\mu \;\;C}^{A}\varphi ^{C}\right) \left( 
\mathrm{D}_{\;\;\;D}^{\mu B}\varphi ^{D}\right) -V\left( \varphi ^{A}\right)
\right) ,  \label{m68}
\end{equation}
invariant under the deformed gauge transformations 
\begin{equation}
\bar{\delta}_{\epsilon }A_{\mu }^{a}=\left( \mathrm{D}_{\mu }\right)
_{\;\;b}^{a}\epsilon ^{b},\;\bar{\delta}_{\epsilon }\varphi ^{A}=\mathrm{g}%
T_{a\;\;B}^{A}\varphi ^{B}\epsilon ^{a}.  \label{m69}
\end{equation}
The modification of the gauge transformations, as well as the appearance of
new such transformations in connection with the matter fields, is
essentially due to the deformation of the secondary first-class constraints
like in (\ref{m65}). If in (\ref{m68}--\ref{m69}) we make the transformation 
$T_{a\;\;B}^{A}\rightarrow \mathrm{i}T_{a\;\;B}^{A}$, we derive the
non-abelian analogue of scalar electrodynamics.

\subsection{Vector fields coupled to Dirac fields}

Finally, we examine the consistent couplings between a set of vector fields
and a collection of massive Dirac fields. In view of this, we start from the
Lagrangian action 
\begin{equation}
\tilde{S}_{0}^{\mathrm{L}}\left[ A_{\mu }^{a},\psi _{\;A}^{\alpha },\bar{\psi%
}_{\alpha }^{\;A}\right] =\int \mathrm{d}^{4}x\left( -\frac{1}{4}F_{\mu \nu
}^{a}F_{a}^{\mu \nu }+\bar{\psi}_{\alpha }^{\;A}\left( \mathrm{i}\left(
\gamma ^{\mu }\right) _{\;\;\beta }^{\alpha }\partial _{\mu }-m\delta
_{\;\;\beta }^{\alpha }\right) \psi _{\;A}^{\beta }\right) ,  \label{m70}
\end{equation}
where $\psi _{\;A}^{\alpha }$ and $\bar{\psi}_{\alpha }^{\;A}$ denote the
fermionic spinor components of the Dirac fields $\psi _{\;A}$ and $\bar{\psi}%
^{\;A}$. Action (\ref{m70}) is invariant under the gauge transformations 
\begin{equation}
\delta _{\epsilon }A_{\mu }^{a}=\partial _{\mu }\epsilon ^{a},\;\delta
_{\epsilon }\psi _{\;A}^{\alpha }=0,\;\delta _{\epsilon }\bar{\psi}_{\alpha
}^{\;A}=0.  \label{m71}
\end{equation}
The purely matter theory is subject to the second-class constraints 
\begin{equation}
\bar{\chi}_{\;A}^{\alpha }\equiv \bar{\Pi}_{\;A}^{\alpha }+\frac{\mathrm{i}}{%
2}\left( \gamma ^{0}\right) _{\;\;\beta }^{\alpha }\psi _{\;A}^{\beta
}\approx 0,\;\chi _{\alpha }^{\;A}\equiv \Pi _{\alpha }^{\;A}+\frac{\mathrm{i%
}}{2}\left( \gamma ^{0}\right) _{\;\;\alpha }^{\beta }\bar{\psi}_{\beta
}^{\;A}\approx 0,  \label{m72}
\end{equation}
where $\Pi _{\alpha }^{\;A}$ and $\bar{\Pi}_{\;A}^{\alpha }$ represent the
canonical momenta respectively conjugated to $\psi _{\;A}^{\alpha }$ and $%
\bar{\psi}_{\alpha }^{\;A}$. By eliminating the constraints (\ref{m72}) with
the help of the Dirac bracket, we find for the model under consideration
that 
\begin{equation}
y^{\alpha _{0}}=\left( \psi _{\;A}^{\alpha },\bar{\psi}_{\alpha
}^{\;A}\right) ,  \label{m73}
\end{equation}
where the fundamental Dirac brackets are expressed by 
\begin{equation}
\left[ \psi _{\;A}^{\alpha },\psi _{\;A}^{\beta }\right] =0,\;\left[ \bar{%
\psi}_{\alpha }^{\;A},\bar{\psi}_{\beta }^{\;B}\right] =0,\;\left[ \psi
_{\;A}^{\alpha },\bar{\psi}_{\beta }^{\;B}\right] =-\mathrm{i}\left( \gamma
^{0}\right) _{\;\;\beta }^{\alpha }\delta _{A}^{\;\;B}.  \label{m74}
\end{equation}
The first-class Hamiltonian is of the type (\ref{m3}), with 
\begin{equation}
\bar{h}_{0}=-\bar{\psi}_{\alpha }^{\;A}\left( \mathrm{i}\left( \gamma
^{i}\right) _{\;\;\beta }^{\alpha }\partial _{i}-m\delta _{\;\;\beta
}^{\alpha }\right) \psi _{\;A}^{\beta }.  \label{m75}
\end{equation}
If we make the identifications 
\begin{equation}
y^{\alpha _{1}}\longleftrightarrow \psi _{\;A}^{\alpha },\;z_{\alpha
_{1}}\longleftrightarrow \bar{\psi}_{\alpha }^{\;A},\;\sigma _{\;\;\beta
_{1}}^{\alpha _{1}}\longleftrightarrow -\mathrm{i}\left( \gamma ^{0}\right)
_{\;\;\beta }^{\alpha }\delta _{A}^{\;\;B},  \label{m76}
\end{equation}
from (\ref{m28}) we deduce that the conserved charge $b_{a}$ reads as 
\begin{equation}
b_{a}=\mathrm{i}\bar{\psi}_{\alpha }^{\;B}\left( \gamma ^{0}\right)
_{\;\;\beta }^{\alpha }T_{a\;\;B}^{A}\psi _{\;A}^{\beta }.  \label{m77}
\end{equation}
On account of (\ref{m75}) we derive that 
\begin{equation}
\left[ b_{a},\bar{H}_{0}\right] =-\partial _{i}\left( \mathrm{i}\bar{\psi}%
_{\alpha }^{\;B}\left( \gamma ^{i}\right) _{\;\;\beta }^{\alpha
}T_{a\;\;B}^{A}\psi _{\;A}^{\beta }\right) ,  \label{m78}
\end{equation}
hence the conserved currents will be 
\begin{equation}
j_{a}^{i}=\mathrm{i}\bar{\psi}_{\alpha }^{\;B}\left( \gamma ^{i}\right)
_{\;\;\beta }^{\alpha }T_{a\;\;B}^{A}\psi _{\;A}^{\beta }.  \label{m79}
\end{equation}
In order to fully determine the interacting theory, it remains to analyse
the higher-order deformations of the BRST-invariant Hamiltonian. Direct
computation yields 
\begin{equation}
\left[ j_{b}^{i},b_{a}\right] =-f_{\;\;ab}^{c}j_{c}^{i},  \label{m80}
\end{equation}
such that we are in case a) (see (\ref{m43})). Consequently, we find that $%
H_{2}$ is like in (\ref{m44}), and also $H_{3}=H_{4}=\cdots =0$.

Putting together the results obtained until now, it results that the
deformed BRST charge and new BRST-invariant Hamiltonian can be written as 
\begin{eqnarray}
&&\Omega =\int \mathrm{d}^{3}x\left( \pi _{a}^{0}\eta _{1}^{a}-\left( \left( 
\mathrm{D}_{i}\right) _{a}^{\;\;b}\pi _{b}^{i}-\mathrm{ig}\bar{\psi}_{\alpha
}^{\;B}\left( \gamma ^{0}\right) _{\;\;\beta }^{\alpha }T_{a\;\;B}^{A}\psi
_{\;A}^{\beta }\right) \eta _{2}^{a}+\right.   \nonumber \\
&&\left. \frac{1}{2}\mathrm{g}f_{\;\;ab}^{c}\mathcal{P}_{2c}\eta
_{2}^{a}\eta _{2}^{b}\right) ,  \label{m81}
\end{eqnarray}
\begin{eqnarray}
&&H_{\mathrm{B}}=\int \mathrm{d}^{3}x\left( \bar{\psi}_{\alpha }^{\;B}\left( 
\mathrm{i}\left( \gamma ^{i}\right) _{\;\;\beta }^{\alpha }\mathrm{D}%
_{i\;\;B}^{A}-m\delta _{\;\;\beta }^{\alpha }\delta _{\;\;B}^{A}\right) \psi
_{\;A}^{\beta }+\frac{1}{2}\pi _{ia}\pi _{i}^{a}+\right.   \nonumber \\
&&\frac{1}{4}\tilde{F}_{ij}^{a}\tilde{F}_{a}^{ij}-A_{0}^{a}\left( \left( 
\mathrm{D}_{i}\right) _{a}^{\;\;b}\pi _{b}^{i}-\mathrm{ig}\bar{\psi}_{\alpha
}^{\;B}\left( \gamma ^{0}\right) _{\;\;\beta }^{\alpha }T_{a\;\;B}^{A}\psi
_{\;A}^{\beta }\right) +  \nonumber \\
&&\left. \left( \eta _{1}^{a}-\mathrm{g}f_{\;\;bc}^{a}\eta
_{2}^{b}A_{0}^{c}\right) \mathcal{P}_{2a}\right) .  \label{m82}
\end{eqnarray}
From the analysis of the above quantities, we see that the modified
constraints are the secondary ones, namely, 
\begin{equation}
\gamma _{2a}\equiv -\left( \left( \mathrm{D}_{i}\right) _{a}^{\;\;b}\pi
_{b}^{i}-\mathrm{ig}\bar{\psi}_{\alpha }^{\;B}\left( \gamma ^{0}\right)
_{\;\;\beta }^{\alpha }T_{a\;\;B}^{A}\psi _{\;A}^{\beta }\right) \approx 0,
\label{m83}
\end{equation}
and they form again a Lie algebra in terms of the Dirac bracket, just like
in (\ref{m32}). The antighost number zero piece from (\ref{m82}) 
\begin{eqnarray}
&&\tilde{H}_{0}=\int \mathrm{d}^{3}x\left( \bar{\psi}_{\alpha }^{\;B}\left( 
\mathrm{i}\left( \gamma ^{i}\right) _{\;\;\beta }^{\alpha }\mathrm{D}%
_{i\;\;B}^{A}-m\delta _{\;\;\beta }^{\alpha }\delta _{\;\;B}^{A}\right) \psi
_{\;A}^{\beta }+\frac{1}{2}\pi _{ia}\pi _{i}^{a}+\right.   \nonumber \\
&&\left. \frac{1}{4}\tilde{F}_{ij}^{a}\tilde{F}_{a}^{ij}-A_{0}^{a}\left(
\left( \mathrm{D}_{i}\right) _{a}^{\;\;b}\pi _{b}^{i}-\mathrm{ig}\bar{\psi}%
_{\alpha }^{\;B}\left( \gamma ^{0}\right) _{\;\;\beta }^{\alpha
}T_{a\;\;B}^{A}\psi _{\;A}^{\beta }\right) \right) ,  \label{m84}
\end{eqnarray}
gives the deformed first-class Hamiltonian, while the term from (\ref{m82})
linear in the antighosts indicates that the Dirac brackets between the
constraints and the first-class Hamiltonian change like in (\ref{m67}). In
conclusion, we are led to the Hamiltonian formulation of the model
describing the interaction between Yang-Mills fields and a collection of
spinor fields.

The Lagrangian setting of this interacting model is described by the action 
\begin{equation}
S^{\mathrm{L}}\left[ A_{\mu }^{a},\varphi ^{A}\right] =\int \mathrm{d}%
^{4}x\left( -\frac{1}{4}\tilde{F}_{\mu \nu }^{a}\tilde{F}_{a}^{\mu \nu }+%
\bar{\psi}_{\alpha }^{\;B}\left( \mathrm{i}\left( \gamma ^{\mu }\right)
_{\;\;\beta }^{\alpha }\mathrm{D}_{\mu \;\;B}^{A}-m\delta _{\;\;\beta
}^{\alpha }\delta _{\;\;B}^{A}\right) \psi _{\;A}^{\beta }\right) ,
\label{m85}
\end{equation}
subject to the deformed gauge transformations 
\begin{equation}
\bar{\delta}_{\epsilon }A_{\mu }^{a}=\left( \mathrm{D}_{\mu }\right)
_{\;\;b}^{a}\epsilon ^{b},\;\bar{\delta}_{\epsilon }\psi _{\;A}^{\alpha }=%
\mathrm{g}T_{a\;\;A}^{B}\psi _{\;B}^{\alpha }\epsilon ^{a},\;\bar{\delta}%
_{\epsilon }\bar{\psi}_{\alpha }^{\;A}=-\mathrm{g}T_{a\;\;B}^{A}\bar{\psi}%
_{\alpha }^{\;B}\epsilon ^{a},  \label{m86}
\end{equation}
which are again a consequence of the new constraints (\ref{m83}). Like in
the scalar case, if in (\ref{m85}--\ref{m86}) we perform the replacement $%
T_{a\;\;B}^{A}\rightarrow \mathrm{i}T_{a\;\;B}^{A}$ and consider an $SU(3)$
gauge algebra, we arrive at quantum chromodynamics.

\section{Conclusion}

In conclusion, in this paper we have investigated the consistent Hamiltonian
interactions that can be introduced between a set of vector fields and a
system of matter fields by using some cohomological techniques. This problem
involves two steps. Initially, we deform the ``free'' BRST charge, and
subsequently approach the deformation of the BRST-invariant Hamiltonian.
Related to the BRST charge, we notice that only the first-order deformation
is non-trivial, while its consistency requires the deformed first-class
constraints form a Lie algebra. Regarding the BRST-invariant Hamiltonian, we
have shown that the first-order interaction contains two terms. The first
one describes an interaction among the vector fields. The second term
appears only if the matter theory possesses some conserved Hamiltonian
currents, and is of the form $j_{a}^{\mu }A_{\mu }^{a}$, where $j_{a}^{\mu }$
are the currents. The second-order deformation of the BRST-invariant
Hamiltonian contains the spatial part of the quartic vertex of pure
Yang-Mills theory. If the currents $j_{a}^{i}$ transform under the deformed
gauge transformations according to the adjoint representation of the Lie
gauge algebra, then all the other deformations involving the matter fields,
of order two and higher, vanish. In the opposite case, at least the
second-order deformation implying matter fields is non-vanishing, but in
principle there might be other non-trivial terms. The general procedure has
been applied to the study of the interactions between a set of vector fields
and a collection of real scalar fields, respectively, a set of Dirac fields.

\section*{Acknowledgment}

This work has been supported by a Romanian National Council for Academic
Scientific Research (CNCSIS) grant.

\end{document}